# D-band MUTC Photodiode Module for Ultra-Wideband 160 Gbps Photonics-Assisted Fiber-THz Integrated Communication System

Yuxin Tian,[1,†] Yaxuan Li,[2,†] Bing Xiong,[1,*] Junwen Zhang,[2,*] Changzheng Sun,[1] Zhibiao Hao,[1] Jian Wang,[1] Lai Wang,[1] Yanjun Han,[1] Hongtao Li,[1] Lin Gan,[1] Nan Chi,[2] and Yi Luo[1]

[1] Beijing National Research Center for Information Science and Technology (BNRist)/ Department of Electronic Engineering, Tsinghua University, Beijing 100084, China
[2] Key Laboratory of Information Science of Electromagnetic Waves (MoE), Department of Communication Science and Engineering, Fudan University, Shanghai 200433, China
† These authors contributed equally to this work
* Corresponding authors: bxiong@tsinghua.edu.cn; junwenzhang@fudan.edu.cn

**Abstract:** Current wireless communication systems are increasingly constrained by insufficient bandwidth and limited power output, impeding the achievement of ultra-high-speed data transmission. The terahertz (THz) range offers greater bandwidth, but it also imposes higher requirements on broadband and high-power devices. In this work, we present a modified uni-traveling-carrier photodiode (MUTC-PD) module with WR-6 waveguide output for photonics-assisted fiber-THz integrated wireless communications. Through the optimization of the epitaxial structure and high-impedance coplanar waveguide (CPW), the fabricated 6-μm-diameter MUTC-PD achieves a high output power of −0.96 dBm at 150 GHz and ultra-flat frequency response at D-band. The MUTC-PD is subsequently packaged into a compact WR-6 module, incorporating planar-circuit-based RF-choke, DC-block and probe. The packaged PD module demonstrates high saturation power and flat frequency responses with minimal power roll-off of only ~2 dB over 110-170 GHz. By incorporating the PD module into a fiber-THz integrated communication system, high data rate of up to 160 Gbps with 16 quadrature amplitude modulation (QAM) has been successfully secured. The demonstration verifies the potential of the PD module for ultra-broadband and ultra-high-speed THz communications, setting a foundation for future research in high-speed data transmission.

## 1. Introduction

Radio-over-fiber (RoF) technology is a cutting-edge communication approach that enables seamless integration of the optical and wireless network, combining the benefits of both radio and optical communication systems.[1-5] By leveraging the wide bandwidth, low optical loss and electromagnetic interference immunity of optical fibers, RoF technology serves as an attractive solution for wireless access to high-speed data communications. With the increasing number of wireless end users and the rapid development of multimedia services, such as virtual and augmented reality (VAR), ultra-high definition (UHD) video, and e-commerce, the carrier frequency in 6G is prompted to the THz domain (0.1-10 THz) to address the extensive use of microwave bands and lack of wireless communication bandwidth.[6-10] In photonics-assisted fiber-THz integrated communication system, photodiode (PD) as a key optical-to-electrical (O/E) conversion device is capable of generating THz signals in the downlink. Figure 1 depicts potential application scenarios for high bit-rate, high-capacity THz wireless communications enabled by high-performance PDs, facilitating delay-free human-computer interaction and rapid downloading of large volumes of data, ultimately enhancing convenience and intelligence in daily life. Particularly in scenarios with elevated data rate requirements, such as high-quality three-dimensional (3D) video and the next generation of VAR, the photonic-based THz generation method facilitates more efficient utilization of bandwidth resources, consequently achieving higher data rate communication.

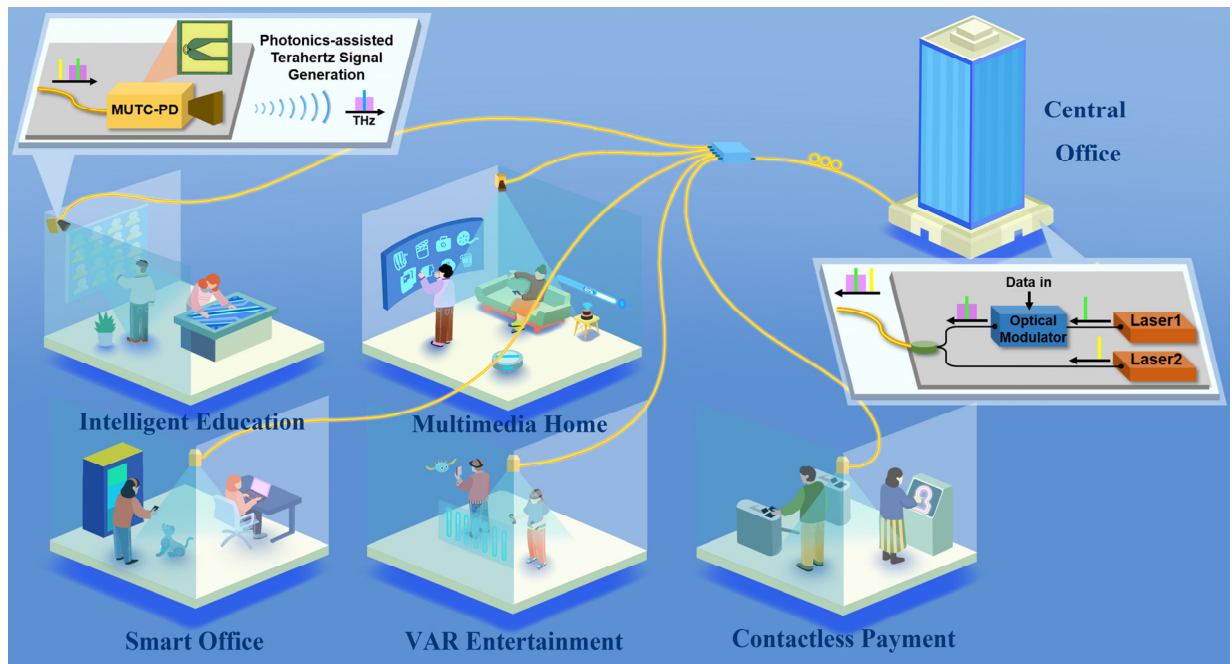

**Figure 1.** Prospective application scenarios of high bit-rate, high-capacity terahertz wireless communications, including intelligent human-computer interactive education, multimedia smart home, smart office, augmented/virtual reality (VAR) entertainment, and smart travel and contactless payment.

To achieve wide bandwidth and high output power, various PD structures have been developed. The saturation power of conventional p-i-n PDs is limited by space-charge effects in the absorption region.[11] Since the uni-traveling carrier photodiode (UTC-PD) was first demonstrated by NTT in 1997, it has been widely employed to simultaneously achieve a high 3-dB bandwidth and a high-saturation output power.[12-17] The D-band is commonly used for THz wireless communications because of its broad bandwidth with low atmospheric attenuation. For D-band applications, modified uni-traveling carrier photodiode (MUTC-PD) is proposed to further improve the space charge tolerance by adding a cliff layer.[18-20] A flip-chip bonded 6-μm-diameter MUTC-PD with 100 GHz bandwidth attains a maximum output power of over −10 dBm at 160 GHz. PDs with diameters larger than 8-μm are able to deliver output power over −4 dBm (saturation current >23 mA), but at the expense of reduced bandwidth.[21] To reduce the roll-off at high frequencies, a short section of low-inductance transmission line on AlN submount is employed. A 4-μm-diameter PD exhibits a 3-dB bandwidth of 145 GHz, with frequency response roll-off of only 6 dB up to 170 GHz. A 9-μm-diameter PD with 75 GHz bandwidth reveals a maximum RF output power of −2.6 dBm at 160 GHz, and a saturation current of 40 mA.[22] Another MUTC-PD with optimized epitaxy structure is flip-chip bonded on diamond submount. Devices with 6-μm and 4-μm diameter exhibit 3-dB bandwidth of 129 GHz and 150 GHz, respectively. The output power and photocurrent of the 6-μm PD reach 1.66 dBm and 11 mA at 130 GHz, while that for the 4-μm diameter PD are −3 dBm and 8 mA at 150 GHz.[23] By inserting an additional p-doped layer in the depletion region, near-ballistic transport and high output photocurrent is achieved.[24-26] The near ballistic UTC-PD (NBUTC-PD) with an active area of 28 μm$^2$ demonstrates a 3-dB bandwidth >110 GHz and a saturation current of 7.4 mA at 110 GHz when flip-chip bonded onto an AlN-based pedestal. By inserting an additional center bonding pad, the PD exhibits a higher saturation photocurrent of 13.6 mA and output power of 1.85 dBm under −3 V bias voltage.[27] However, the bandwidth of the above PDs is insufficient to fulfill the need for D-band applications. Furthermore, the PDs are flip-chip bonded onto AlN/diamond submount for improved output power, which increases the fabrication complexity. In our previous work, a 4.5-μm MUTC-PD exhibits 3-dB bandwidth of 150 GHz with output power of 1.72 dBm at 140 GHz, but its power roll-off reaches 3 dB in the D-band.[28] A waveguide PD with an active area of 5 × 6 μm$^2$ has demonstrated ultrawide 3-dB bandwidth of 153 GHz, but the saturation current is only 7 mA with RF power limited to −5.6 dBm at 130 GHz.[29] Therefore, it is necessary to extend the bandwidth of the PD to ensure flat frequency response over the entire D-band (110-170 GHz). Meanwhile, it is crucial to enhance the output power of the PD, so that the power amplifier in the wireless communication system can be removed, thus reducing the system complexity and costs.

In order to facilitate application in more scenarios, the PD chip needs to be packaged into a compact module. In Ref[30], a photodetector module with a state-of-the-art 0.8 mm RF connector supports 145 GHz transmission bandwidth. The photodiode module exhibits frequency response with less than 3-dB roll-off within the 145 GHz bandwidth. For system applications at frequencies over 145 GHz, PD modules with standard waveguide outputs are desirable. A pin-PD is packaged with standard WR-10 waveguide output for operation in the W-band, but the maximum mm-wave output power is limited to −10 dBm at 100 GHz.[31] A commercialized UTC-PD module with WR-6 output port achieves output power around −7 dBm over the entire D-band.[32] The enhanced output power is realized by integrating resonant matching circuits with the UTC-PD on InP substrate[33] However, currently there are few reports on the packaging design and fabrication of PD modules, especially for high frequencies at D-band.

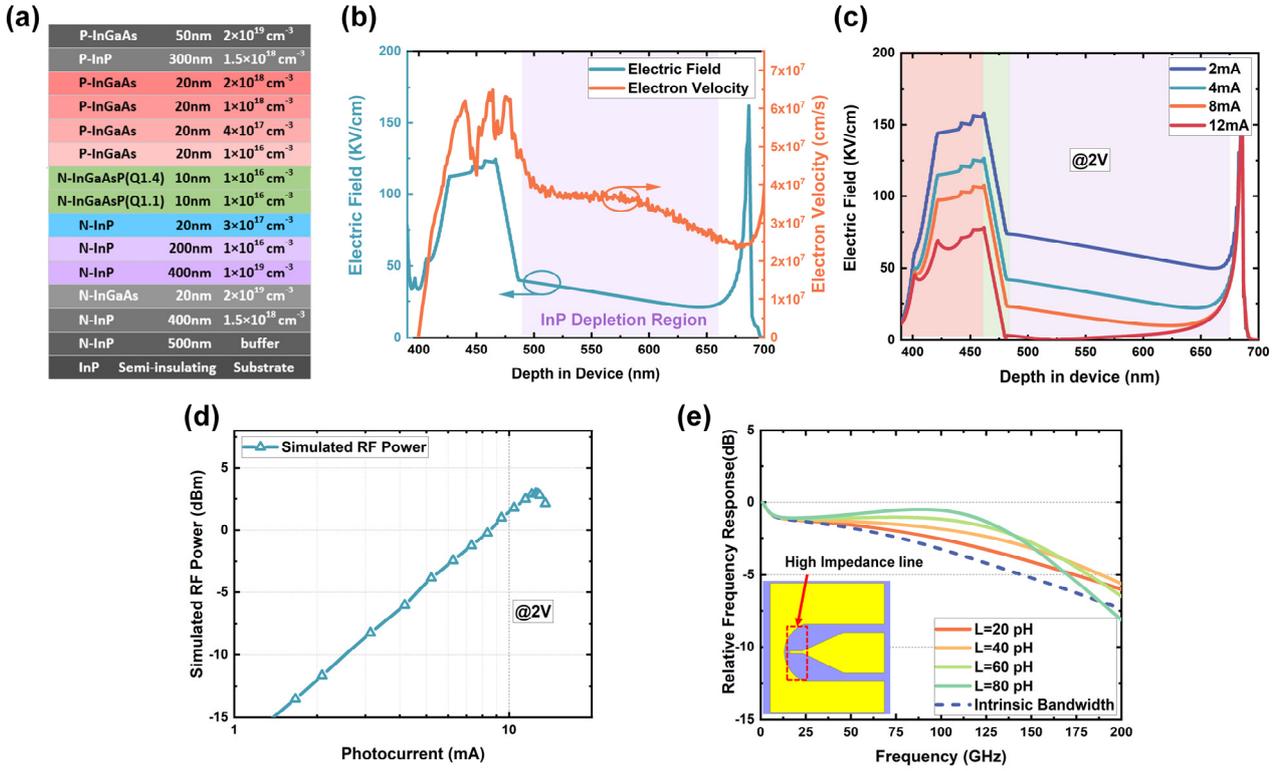

**Figure 2.** Design and performance of MUTC-PD. a) Epitaxy structure of the proposed MUTC-PD. b) Electric field and electron velocity distributions in the depletion region of the PD. c) Electric field distributions in the depletion region at different photocurrents. d) Simulated saturation performance of the PD. e) Variation of the PD frequency response with inductance introduced by the high-impedance CPW electrodes.

There are some recent demonstrations of D-band photonics-assisted THz communication systems using PD for photomixing at the transmitter end and using envelope detector (ED) or mixers for heterodyne reception at the receiver end.[34-39] Limited by available bandwidth and spectrum efficiency, the THz communication data rates achieved in these demonstrations were less than 100 Gbps, and the maximum baud rates of high-order modulation format such as 16 quadrature amplitude modulation (QAM) signals did not exceed 20 Gbaud. Certainly, there are other multiple-input multiple-output (MIMO) demonstrations in the D-band that have achieved ultra-high data rates. By employing truncated probabilistic shaping (PS) and MIMO Volterra compensation, delivery of 103.2 Gb/s 4096 QAM signal has been implemented.[40] Total data rate up to 352 Gb/s is experimentally achieved in a single 2×2 multiuser MIMO system.[41] The employment of PS, Nyquist-shaping, 4×4 MIMO technology and look-up-table improved the transmission capacity to 1.056 Tb/s.[42] However, the application of MIMO and other technologies have also brought about significant increase in link costs and digital signal processing (DSP) complexity.

In this paper, a D-band MUTC-PD module is employed in photonics-assisted fiber-THz integrated wireless communications to achieve the highest single channel data rate. The broadband PD module is achieved by flat PD chip response, low-loss transition circuits as well as optimized packaging process. The PD chip with optimized epitaxy structure and specially designed coplanar waveguide (CPW) electrodes demonstrates both flat frequency response and high output power at D-band. By employing low-loss bias-tee and probe, the WR-6 PD module exhibits flat output power with roll-off of ~2 dB under various photocurrents and high saturation power. The packaged PD module is employed in a D-band wireless communication system, transmitting signals of various modulation formats. Data rates up to 160 Gbps for 16 QAM signals, 135 Gbps for 8QAM signals are achieved over a 10-km standard single-mode fiber (SMF) and a 1-m wireless transmission. A maximum symbol rate of 60 Gbaud quadrature phase shift keying (QPSK) signal has been transmitted successfully, achieving a data rate of 120 Gbps. Additionally, in a 1-m wireless back-to-back (B2B) transmission, a data rate of 100 Gbps is recorded with 20 Gbaud 32QAM signals. The experimental results fully confirm the broadband characteristics of the PD and its potential in high-speed THz wireless communications.

## 2. Broadband MUTC-PD Chip

To effectively estimate the 3-dB bandwidth of the MUTC-PD and investigate its main limiting factors, a two-port equivalent circuit model[43] is adopted, which takes both transit time and resistance-capacitance (RC) delay into consideration. A 6-μm-diameter PD is employed to extend the RC-limited bandwidth while ensuring high output power. In addition, the effect of CPW electrodes on the frequency response is simulated using the circuit model to ensure flat response for the D-band.

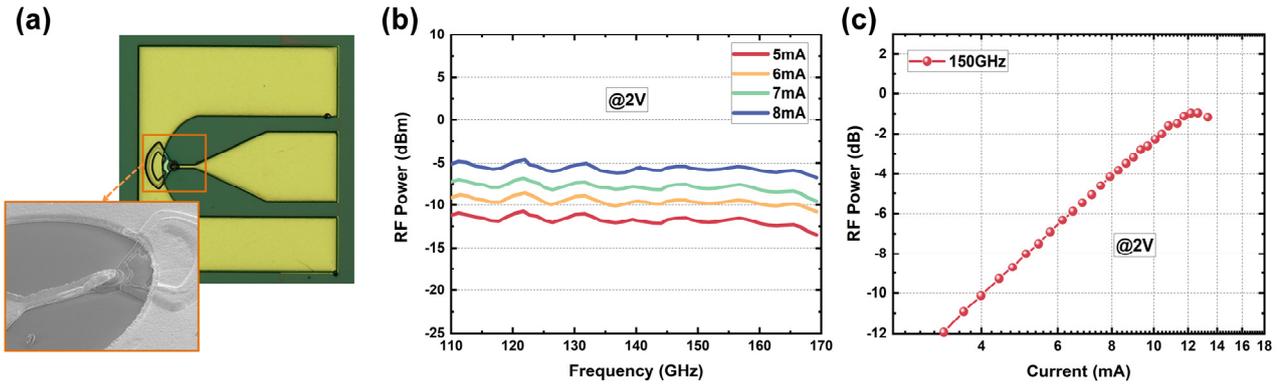

**Figure 3.** a) Microscope and SEM images of the PD. b) Frequency responses and c) saturation characteristics of the 6-μm PD under −2 V bias voltage.

The epitaxy layer structure of the back-illuminated MUTC-PD is demonstrated in Figure 2a. The 80 nm-thick gradient p-doped (from $1\times10^{16}$ cm$^{-3}$ to $2\times10^{18}$ cm$^{-3}$) absorption layer is employed to create a quasi-electric field to facilitate electron drift into the depleted collection layer. The 20 nm-thick n-doped InP cliff layer is adopted to adjust the electric field within the absorption and collection regions. An optimized doping concentration of $3\times10^{17}$ cm$^{-3}$ is employed to finetune the internal electric field for accelerated electron transportation.[43] As plotted in Figure 2b, the electric field in the InP-depletion region is 20-40 kV/cm, corresponding to an electron velocity of $3.5\times10^7$ cm/s, which is in the range of electron velocity overshoot.[44] The cliff layer also helps improve the high saturation performance of the PD. The n-doped cliff layer can enhance the electric field within the depleted absorption layer, thereby suppressing the space-charge effect. Meanwhile, it also tunes the electric field in the InP depletion layer to ensure electron transport at overshoot velocity and avoid electron accumulation. The electric field distributions of the proposed epitaxy structure under different photocurrents are plotted in Figure 2c. As the optical power increases, the electric field is reduced due to the screening of photo-generated carriers. When the photocurrent reaches 12 mA, the electric field in depletion region is reduced to zero because of the space charge effect. In comparison, the electric field in the absorption region of a traditional UTC-PD collapses at much lower input optical power.[43] A physically-based simulator is employed to investigate the saturation performance of the PD.[45, 46] The simulated saturation characteristics of the PD at 150 GHz under 2 V reverse bias is plotted in Figure 2d, indicating a saturation photocurrent of 12.5 mA. It is worth noting that the simulation only considers the transit-time-limited bandwidth and fails to take the RC-limited bandwidth into account, thus resulting in an over estimation of the output power.

Additionally, inductive peaking is implemented with a high impedance section in the CPW electrodes, which helps achieve improved impedance matching for the capacitive PD. By adjusting the inductance, a flat frequency response with reduced power roll-off can be achieved at the high frequencies of interest. Figure 2e shows the relative frequency responses of the 6-μm-diameter PD with different high-impedance CPWs. It is evident that a small inductance leads to a slight increase in the output power at the D-band, while a large inductance results in steep roll-off at higher frequencies. Therefore, a moderate inductance of 40 pH is employed to secure high output power and flat frequency response simultaneously. The 3-dB bandwidth of the proposed PD is estimated to be 147 GHz, and the frequency response exhibits a roll-off of less than 2 dB in the range of 110-170 GHz.

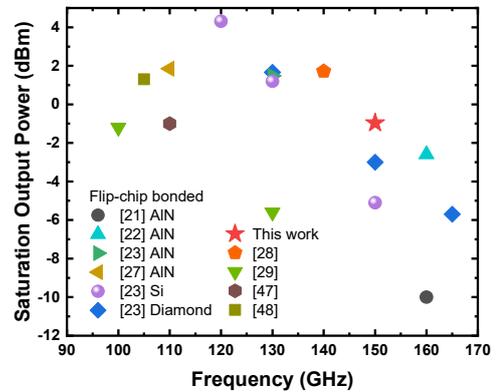

**Figure 4.** Comparison of saturation output power for UTC-PDs reported in the literature.

The microscope photo of the fabricated 6-μm-diameter PD is illustrated in Figure 3a, and the detailed mesa structure is captured by scanning electron microscope (SEM). The frequency responses of the 6-μm-diameter PD at different photocurrent levels measured by two-laser heterodyne system are plotted in Figure 3b. Thanks to the high-speed electron transport and the inductive CPW design, the PD chip exhibits ultra-flat frequency response with a roll-off of only ~1.6 dB over the entire D-band. Thanks to the optimized cliff layer design, which makes the PD less susceptible to the space charge effect, the saturation power at 150 GHz is −0.96 dBm, corresponding to a photocurrent of 12.6 mA, as shown in Figure 3c. The saturated power is measured at 150 GHz, which corresponds to the 3-dB bandwidth of the device, indicating a 3 dB drop in output power with respect to low frequencies. Figure 4 gives an overview of the saturation performance of D-band PDs reported in the literature. In Refs. [21-23, 27], the PD chips are flip-chip bonded onto submount with high thermal conductivity to enhance the saturation output power. Thanks to the optimized epitaxy structure and high-impedance CPW design, our PD exhibits flat response together with high output power without flip-chip bonding onto a heatsink.

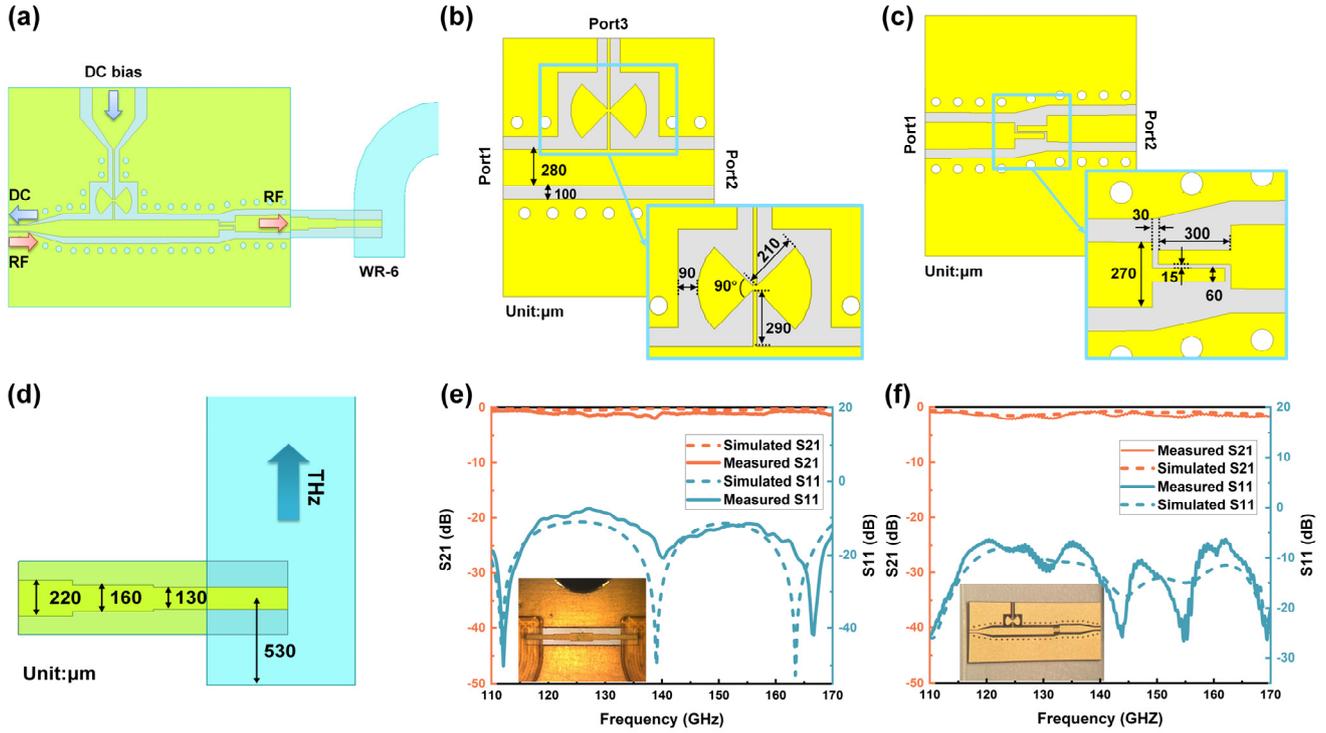

**Figure 5.** Design and measurement of planar-circuit-based RF-choke, DC-block and probe. a) Top view of the transition circuits. Geometries of the b) shunt-connected radial-stubs based RF-choke, c) coupled-line DC-block and d) stepped-impedance probe. Measured and simulated results of the back-to-back e) probe package and f) bias-tee.

## 3. MUTC-PD Module with WR-6 Waveguide Output

### 3.1 Low-loss Transition Circuits

Optimum operation of PD requires a certain reverse DC-bias to build up the internal electric field,[15-17, 43] thus it is essential to incorporate an integrated bias-tee in the PD module. The bias-tee circuit, consisting of an RF-choke and a DC-block, is intended to provide the DC bias and extract the RF signal from the PD chip. Furthermore, a probe is employed to smoothly guide the THz signal from the bias-tee into WR-6 waveguide. The bias-tee and the probe are fabricated on 127-μm-thick quartz substrate for its low dielectric loss. Grounded coplanar waveguide (GCPW) is adopted to confine the electromagnetic field in the laminate, leading to improved coupling efficiency. Metal-filled via holes are employed to connect the top and bottom ground metals, so as to suppress the slot line mode. The top view of the overall transition circuits is illustrated in Figure 5a. Compared with our previous work in the F-band[49] and the G-band[50], the transition circuits for our PD module operating in the D-band faces a challenge to maintain both low loss and resonance-free over a wide frequency range of 60 GHz. In particular, a flat output over 110-170 GHz is essential to support the maximum symbol transmission rates for high-performance wireless communications.

In addition to ensure low-loss transmission at D-band, the RF-choke should provide good isolation for the RF signal. The performance of conventional straight stubs degrades at higher frequencies due to the excitation of higher order modes when the transmission-line width becomes a fraction of the wavelength.[51-53] In order to overcome this problem, radial-line stubs can be employed instead. As shown in Figure 5b, a signal line width of 280 μm with a gap of 100 μm is adopted to ensure an impedance of about 50 Ω at D band and match the impedance of the CPW electrodes on the PD chip. A quarter-wave high-impedance line is connected to a pair of shunt-connected radial stubs to form RF open circuit.[54] The geometry of the proposed RF-choke shown in Figure 5b has been optimized by full-wave 3D simulation based on finite element method (FEM). To achieve low transmission loss, a radial stub with radius of 210 μm and angle of 90° is employed. The length and the gap of the quarter-wave high-impedance line are optimized to be 290 μm and 90 μm, respectively, so as to ensure low insertion loss and high RF-to-DC isolation over 110-170 GHz. The simulated RF-to-DC isolation is at least 22 dB in the D-band.

Quarter-wavelength coupled transmission lines are widely used in directional couplers, filters, and impedance transformers.[55, 56] We adopt the coupled line (CL) as the DC-block, which blocks the DC current, while allows the RF signal to flow through. The geometrical parameters shown in Figure 5c is designed based on FEM for low-loss transmission of the THz signal. The length and width of the quarter-wavelength CL are optimized to be 300 μm and 60 μm, respectively, to secure low-loss transmission in the entire D-band. A gap of 15 μm is chosen to maintain strong coupling. Impedance matching of the CL is crucial for maximum power transmission of THz signals. The even- and odd-mode impedances of the CL are calculated to be 155 Ω and 51 Ω, respectively, resulting in an input/output impedance of 52 Ω.[57] A tapered transition is adopted to achieve smooth impedance transformation from the 50 Ω GCPW to the 52 Ω CL.

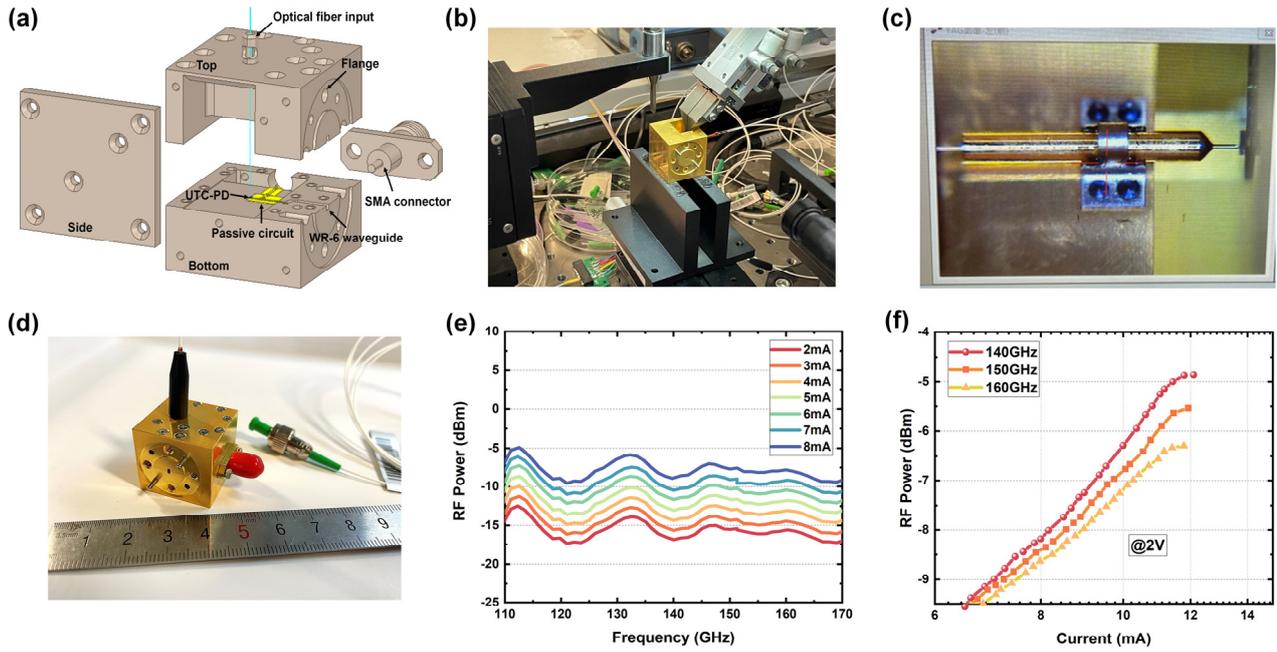

**Figure 6.** Package and performance of MUTC-PD module. a) Internal layout of the PD module with WR-6 output. b) Fiber coupling adjustment by fiber fixture. c) Lensed fiber fixed by laser welding. d) Photograph of the packaged WR-6 PD module. e) Frequency response versus photocurrent at 2 V reverse bias. f) The saturation performance of the PD module under different frequencies.

Stepped-impedance probe is adopted to smoothly transform the impedance from the 50 Ω GCPW to the WR-6 waveguide at the center frequency of 140 GHz. The geometry and the location of the E-plane probe shown in Figure 5d are optimized by full 3-D electro-magnetic simulations. An impedance of 50 Ω can be secured with a microstrip signal line width of 220 μm. As the signal width decreases in steps of 220, 160, and 130 μm, the impedance is gradually increased to match the WR-6 waveguide. In order to ensure maximum coupling efficiency, the optimized location for the probe is found to be 530 μm from the short-end.

For validating the proposed design, back-to-back probe package and bias-tee are fabricated. The measurement and simulation results are plotted against each other in Figures 5e and 5f. The back-to-back probe exhibits a flat transmission. The return loss (RL) is better than 10 dB over most part of the D-band and the insertion (IL) is less than 1.9 dB, corresponding to an IL of 0.8 dB for a single microstrip-to-waveguide transition. The bias-tee performance with a back-to-back configuration demonstrates an average IL around 1.4 dB. The measurement results of both transition circuits are in fair agreement with the simulations indicated by dashed lines.

### 3.2 Packaging and Measurements

In our packaging scheme, the PD chip and the transition circuits are optimized and designed individually to improve the overall performance of the optoelectronic mixing module, resulting in higher design flexibility.[58-60] As demonstrated in Figure 6a, the passive circuits including the probe and the bias-tee are mounted onto the brass carrier by silver-filled epoxy. The passive transitions are connected by wedge bonding with gold wires. The misalignment between the bias-tee and the probe as well as the length and the height of bonding wires are kept at a minimum to suppress the parasitic inductance.[60, 61] Another 6-μm-diameter PD on the same wafer is then flip-chip bonded onto the quartz substrate equipped with 5-μm thick solder bumps.

Fiber coupling is the last step in the packaging of the PD module. A lensed fiber with a spot diameter of 5±1 μm is adopted to couple light to the PD. Active alignment is employed to ensure optimum coupling efficiency. During the fiber coupling, input light with a power of 1 mW is fed into the lensed fiber, and a reverse bias voltage of −1 V is applied to the PD chip via an SMA connector. The photocurrent is monitored as the fiber is adjusted by a fiber fixture, as illustrated in Figure 6b. Once the maximum photocurrent is secured, the optical fiber is fixed by laser welding, as shown in Figure 6c. As laser welding tends to cause a slight shift of the fiber, the responsivity of the packaged PD module is around 0.04 A/W, somewhat lower than that of the PD chip (0.07 A/W).

The photo of the packaged WR-6 PD module is shown in Figure 6d. Optical heterodyne system is used to measure the frequency responses and the saturation performance. The frequency responses tested under different photocurrents with a fixed reverse bias of 2 V are plotted in Figure 6e, revealing flat output RF power over the entire D-band. The power roll-off is only ~2 dB from 110 GHz to 170 GHz under various photocurrents. The over 60 GHz bandwidth of the PD module promises broadband applications at D-band. The saturation behavior measured from 140 GHz to 160 GHz is shown in

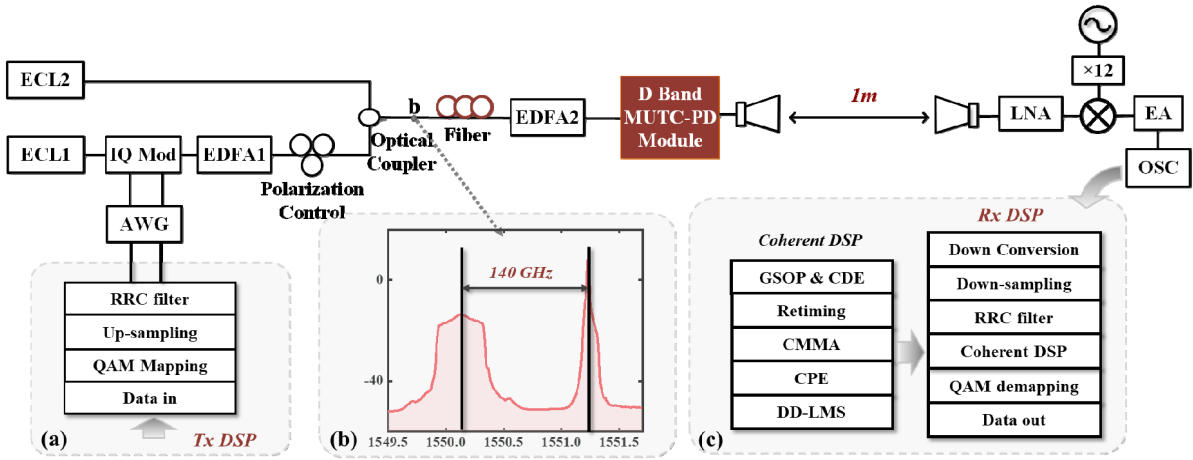

**Figure 7.** The experimental setup of the D-band fiber-THz integrated communication system using the PD module. ECL: external cavity laser, AWG: arbitrary waveform generator, IQ Mod: IQ modulator, EDFA: Erbium doped fiber amplifier, LNA: low noise amplifier, EA: electrical amplifier, OSC: oscilloscope. a) DSP of signal generation on the transmitter side. b) The measured optical spectrum after the optical coupler. c) DSP of signal recovery on the receiver side.

Figure 6f. The photocurrent at 1-dB compression is very much similar, which is 12.1, 11.9, and 11.7 mA under 140 GHz, 150 GHz, 160 GHz, corresponding to an output power of −4.8, −5.5, and −6.2 dBm, respectively. As a result of the losses caused by the transition circuits and introduced during the packaging process, the saturation power is slightly lower than that of the PD chip. Thanks to the optimized epitaxy structure and CPW electrodes of the MUTC-PD chip, as well as the low-loss passive circuits and optimized packaging process, both wide bandwidth and flat response are achieved. Compared with commercialized UTC-Photomixer module[62], our PD module exhibits a fluctuation of only ±2 dB in the frequency response over 110-170 GHz. Therefore, it can support higher baud rates and achieve higher data rates than previously reported photonic-assisted THz communication systems operating in the D band.[34-39]

## 4. High-bit-rate THz Wireless Communication Experiments

### 4.1 Experiment Setup

The MUTC-PD module has great potential in photonics-assisted THz communication systems, as its flat response over the D band allows a large signal bandwidth, which is suitable for various application scenarios. Experiments are carried out to demonstrate a photonics-assisted high-speed fiber-THz integrated communication system based on the broadband MUTC-PD module without electric amplifier.

Figure 7 shows the experimental setup of the D-band fiber-THz integrated communication system. At the transmitter end, the baseband signal generated by a 120 GSa arbitrary wave generator (AWG) is modulated via an IQ modulator onto the optical carrier at 193.24 THz emitted by an external cavity laser (ECL). After amplification by an erbium doped fiber amplifier (EDFA) and polarization control, it is combined with a local oscillator light at 193.1 THz emitted by ECL-2. The optical spectrum measured after the optical coupler (OC) is shown in Figure 6b. After passing through 10 km SMF and amplification by another EDFA, the signal undergoes photoelectric conversion via the PD module. Then, the signal with a center frequency of 140 GHz is radiated into free space through a horn antenna and transmitted over a wireless distance of 1 m before being received by another horn antenna. The back-to-back (B2B) transmission is achieved by removing the 10 km SMF and only has 1-m wireless distance. The received THz signal first passes through a 19 dB-gain low-noise amplifier (LNA) at D-band, and then enters a mixer for heterodyne reception via frequency down-conversion. After passing through a 55 GHz-bandwidth, 23 dB-gain electric amplifier (EA), it is sampled by a 256 GSa digital storage oscilloscope (OSC) and enters offline DSP. The system bandwidth is limited by the receiving mixer, which has a bandwidth of only 35 GHz over 125-160 GHz. The free-space path loss of a wireless link can be expressed as $A(d,f) = 20\log_{10}(4\pi df/c)$, where $d$ represents the wireless distance, $c$ the speed of light, and $f$ the frequency of the signal[63]. Table 1 summarizes the parameters of key devices adopted in the system, as well as the estimated free-space path losses after transmission through a distance of 1 m over the entire D band.

**Table 1.** Parameters of key devices in the wireless link

| Component | Key parameters |
|---|---|
| HAs (HD-1400SGAH25) | Gain: 25dBi<br>Operating frequency: 113~173 GHz@3dB |
| LNA (TMLA-110170-1940-06) | NF: 6 dB    Gain: 19 dB<br>Operating frequency: 110~170 GHz @3dB |
| Mixer and frequency multiplier (AT-12MIX-125160D1) | RF Frequency Range: 125~160 GHz @3dB<br>LO Multiplier Factor: 6<br>Mixer Type: Sub-Harmonic Mixer |
| EA (SHF S807 C) | NF: 6 dB    Gain: 23 dB<br>Operating frequency: 50 kHz~55 GHz @3dB |
| Free-space path loss | 73.26 dB@110GHz<br>74.02 dB@120GHz<br>74.72 dB@130GHz<br>75.36 dB@140GHz<br>75.96 dB@150GHz<br>76.52 dB@160GHz<br>77.05 dB@170GHz |

Figures 7a and 7c show the DSP for signal generation on the transmitter side and signal recovery on the receiver side, respectively. In the offline DSP on the receiver side, the intermediate frequency (IF) signal is first down-converted, then down-sampled, and processed through a matched filter. The coherent DSP algorithm is then applied, which mainly includes Gram-Schmidt orthogonalization procedure (GSOP) for IQ imbalance compensation, clock recovery, cascade multimode algorithm (CMMA) equalization, frequency/phase offset correction, decision-directed least mean square (DDLMS) algorithm, and finally symbol-to-bit demapping to calculate the bit error rate. We have transmitted signals in several modulation formats including QPSK, 8QAM, 16QAM, and 32QAM, with a maximum symbol rate of 65 Gbaud.

### 4.2 Results and Discussion

Figure 8 shows the bit-error-rate (BER) curves for 25 Gbaud 16 QAM, 40 Gbaud 8 QAM, and 50 Gbaud QPSK signals, respectively. The soft decision forward-error-correction (SD-FEC) threshold is 2e−2. With 15.8 dBm optical power into the PD (1.5 mA photocurrent), the system can support SMF and B2B transmission of 25 Gbaud 16QAM and 50 Gbaud QPSK signals, achieving a transmission rate of 100 Gbps. For 40 Gbaud 8QAM signals, the optical power needs to be increased to 17 dBm, corresponding to a photocurrent of 2 mA, to enable successful transmission over 10-km SMF. As the photocurrent increases, the BER decreases due to an improved signal-to-noise ratio (SNR). However, excessive photocurrent such as 4 mA leads to nonlinear effect in the receiving amplifier, resulting in an increase in BER. Taking signal-to-noise ratio, nonlinearity, and system power into account, the optical power into the PD is fixed at 18.8 dBm (3 mA photocurrent) for the transmission of signals with different data rates.

Figure 9 illustrates the variation of the BER performance of different modulation formats with the baud rate. The maximum baud rate achieved by each modulation format increases as the order decreases. The 20 Gbaud 32QAM signals achieve a data rate of 100 Gbps in 1-m wireless B2B transmission. The 16QAM signals support transmission with a maximum bandwidth of 40 Gbaud, achieving a highest line rate of 160 Gbps. The 45 Gbaud 8QAM signals achieve a transmission rate of 135 Gbps. As for QPSK, with the high SNR provided by the MUTC-PD module and the effective compensation and equalization of inter-symbol interference (ISI) by DSP, we have achieved faster-than-Nyquist (FTN) transmission of 60-Gbaud QPSK signal despite the uneven frequency response of the overall system. The insets show the received constellations of signals with different modulation formats at the maximum data rate. Figure 10b plots the IF spectrum of the 60 Gbaud QPSK signal. The mixer bandwidth of our system is severely limited, and improved transmission performance is expected by adopting electrical components with wider bandwidth. Nonetheless, the broadband characteristics of the PD has been well validated, and the communication experiment results match the PD test results. Table 2 summarizes some recent demonstrations of D-band photonics-assisted THz communication systems, which mostly adopt standard PAM or QAM formats with single carrier.[34-39] Thanks to its flat response and ultra large bandwidth, the PD module supports transmission of high-order modulation format signals. According to Table 2, by employing high-order modulation formats, this work demonstrates the highest line rate of 160 Gbps, and the corresponding net data rate of 139.13 Gbps also exceeds previous reports. Furthermore, the above results are obtained by employing only simple single carrier modulation and common coherent DSP. In future work, PS, MIMO, complex modulation formats, and advanced DSP can also be adopted in this system to further improve the transmission capacity.

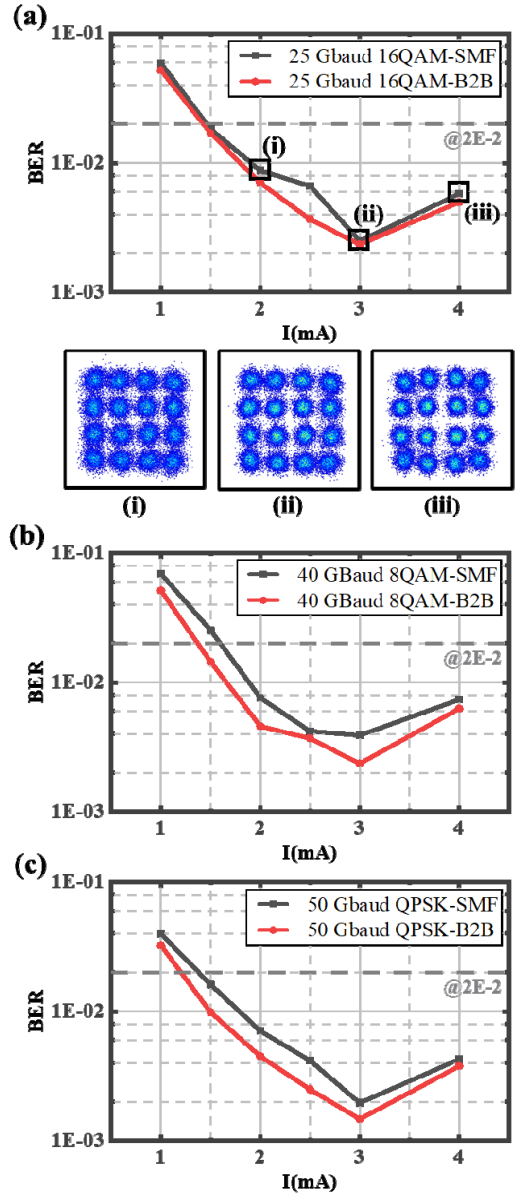

**Figure 8.** BER curves for a) 25 Gbaud 16QAM, b) 40 Gbaud 8QAM, and c) 50 Gbaud QPSK signals, respectively.

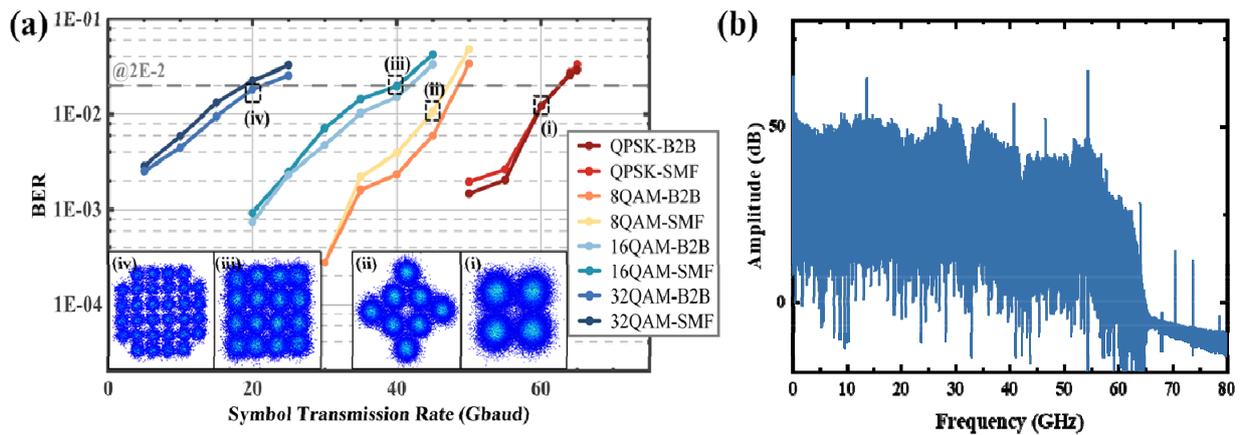

**Figure 9.** a) BER performance of various modulation formats against the baud rate, b) IF spectrum of the 60 Gbaud QPSK signal.

Table 2. Comparison of photonics-assisted D-band THz systems

| Reference | Modulation | Maximum Data Rate / Corresponding Baud Rate of Single Channel Single Polarization | Maximum Net Data Rate of Single Channel Single Polarization | Maximum Baud Rate of Single Channel Single Polarization | Fiber / Wireless Distance |
|---|---|---|---|---|---|
| [34] | QPSK | 92 Gbps/46 Gbaud | 80 Gbps | 46 Gbaud | 80 km/0.6 m |
| [35] | 64QAM | 66 Gbps/11 Gbaud | 61.68 Gbps | 12 Gbaud | 20 km/0 |
| [36] | QPSK | 16 Gbps/8 Gbaud | 12.8 Gbps | 8 Gbaud | 0/4.6 km |
| [37] | 16QAM | 64 Gbps/16 Gbaud | 59.81 Gbps | 16 Gbaud | 25 km/30 m |
| [38] | PAM8 | 60 Gbps/20 Gbaud | 56.07 Gbps | 20 Gbaud | 10 km/3 m |
| [39] | PAM4 | 90 Gbps/45 Gbaud | 84.11 Gbps | 45 Gbaud | 10 km/3 m |
| This work | 16QAM | 160 Gbps/40 Gbaud | 139.13 Gbps | 40 Gbaud | 10 km/1 m |

## 5. Conclusion

In this work, we present a packaged high-speed MUTC-PD at D-band and its application in photonics-assisted fiber-THz integrated wireless communications. By optimizing the epitaxy structure and the CPW electrode, the 6-μm-diameter PD chip exhibits ultra-flat frequency response with a roll-off of only ~1.6 dB over 110-170 GHz. Passive circuits consist of integrated bias-tee and probe are optimized with an insertion loss only 1.5 dB. The packaged PD module demonstrates flat frequency responses and high output power at D-band. In the THz wireless communication experiment, the packaged PD supports the transmission of QPSK, 8QAM, 16QAM, and 32QAM signals beyond 100 Gbps, with the highest data rate of 160 Gbps for 16QAM signal. The high bit-rate performance fully confirms the broadband flat response characteristics of the PD module.

## Supporting Information

Supporting Information is available from the Wiley Online Library or from the author.

## Acknowledgements

This work was supported in part by National Key R&D Program of China (2022YFB2803002); National Natural Science Foundation of China (62235005, 62127814, 62225405, 61975093, 61927811, 61991443, 61925104 and 61974080); and Collaborative Innovation Centre of Solid-State Lighting and Energy-Saving Electronics.

## Data Availability Statement

The data that support the findings of this study are available from the corresponding author upon reasonable request.

## Conflict of Interest